\documentclass{PoS}

\title{Investigation of the spectral lag - energy relation of GRBs registered by INTEGRAL}

\ShortTitle{Spectral lag - energy relation of GRBs}

\author{\speaker{P. Minaev}\\
        Space Research Institute of the Russian Academy of Sciences, Moscow, Russia\\
        E-mail: \email{minaevp@mail.ru}}

\author{A. Pozanenko\\
        Space Research Institute of the Russian Academy of Sciences, Moscow, Russia\\
        E-mail: \email{apozanen@iki.rssi.ru}}

\author{S. Grebenev\\
        Space Research Institute of the Russian Academy of Sciences, Moscow, Russia\\
        E-mail: \email{grebenev@iki.rssi.ru}}

\author{S. Molkov\\
        Space Research Institute of the Russian Academy of Sciences, Moscow, Russia\\
        E-mail: \email{serge.molkov@gmail.com}}

\abstract{We investigated the dependence of spectral lag on energy band based on 28 bright GRBs detected by the SPI and IBIS/ISGRI instruments on the INTEGRAL observatory. It is found that for simple structured bursts or well separated pulses of multi-pulse bursts the spectral lag can be approximated by the relation $ \tau \sim A\lg(E)$, where A is a positive parameter, which correlates with pulse duration. We have not found any negative lag in simple structured bursts or in well separated pulses. While investigating the time profile of the whole burst negative lag may appear due to different spectral parameters of the pulses.}

\FullConference{"An INTEGRAL view of the high-energy sky (the first 10 years)"
9th INTEGRAL Workshop and celebration of the 10th anniversary of the launch,\\
		October 15-19, 2012\\
		Bibliotheque Nationale de France, Paris, France}

\begin{document}

\section{Introduction}
Spectral evolution of gamma-ray bursts (GRB) is one of the most interesting phenomenological properties. In most cases, we observe the evolution from the hard spectrum in the beginning, to soft, in the final phase of gamma-ray burst. Different dependencies between the spectral evolution and other properties of gamma-ray bursts have been found (Hakkila and Preece, 2011).

One of the models describing spectral evolution of GRBs is based on curvature effect of the relativistic shocked shell. At the source, the relativistically expanding shell emits identical pulses from all latitudes. However, when the photons reach the detector, on-axis photons get boosted to higher energy (hard). Meanwhile, off-axis photons get relatively smaller boost and travel longer to reach the detector. Thus, these photons are softer and arrive later than the on-axis photons (Ukwatta et al., 2012).

\section{Instruments}
Data of SPI and IBIS/ISGRI instruments of INTEGRAL observatory were used for analysis. Detailed description of INTEGRAL mission see in (Winkler et al., 2003), detailed description of spectrometer SPI and IBIS/ISGRI experiments see in (Vedrenne et al., 2003) and in (Ubertini et al., 2003), respectively.

These telescopes have large effective area of detectors and good spectral resolution (especially, SPI). But main feature of these instruments is ability to form photon-to-photon light curve - information (energy and time moment of detection) about each registered photon is stored. So data of these instruments are perfect to use in spectral and temporal analysis.

\section{Data analysis and main results}
28 bright GRBs registered by SPI and IBIS/ISGRI experiments were investigated. Well separated pulses of multi-pulse events were investigated independently. So total number of analyzed events was 43.

Method of investigation was based on cross-correlation analysis of light curves in two different energy bands. Details of cross-correlation analysis see in (Band, 1997). In this analysis cross-correlation function (CCF) of two light curves in different energy channels is formed. Position of the maximum in CCF curve determines the value of light curves time offset, which is called spectral lag. Spectral lag is positive when light curve in higher energy band is registered earlier than one in lower energy band.

Algorithm of our analysis consisted of next steps:

1) Building a GRB energy-time diagram (fig.1a) and visual analysis of GRB spectral and temporal properties (hardness of GRB spectrum, number of pulses, duration, etc).

2) Building a light curves in narrow energy bands (up to 25 channels) and selection of time interval for analysis. Time resolution and energy channel width of light curves depend only on GRB properties.

3) Cross-correlation analysis of formed light curves to determine spectral lag between light curve in the first lowest energy channel and light curves in other channels.

4) Approximation of spectral lag - energy relation (fig.1c) using two models (formulas 3.1-3.3), where parameter A (spectral lag index) characterizes spectral evolution.
\begin{eqnarray}
\tau = A\log(E)+B
\label{one}.
\end{eqnarray}

\begin{eqnarray}
 \tau= (A_{1}\log(E)+B_{1})\exp\left[\left[-\frac{\log(E)}{\log(E_{cut})}\right]^{C}\right]+(A_{2}\log(E)+B_{2})\left[1-\exp\left[\left[-\frac{\log(E)}{\log(E_{cut})}\right]^{C}\right]\right]
\label{one}.
\end{eqnarray}
\begin{eqnarray}
B_{2} = E_{cut}(A_{1}-A_{2})+B_{1}
\label{one}.
\end{eqnarray}
In most cases lag - energy relation is well described by simple logarithmic model (3.1) (fig.1c, left). Positive value of the slope of the curve means that light curve in higher energy band is registered earlier than one in lower energy band (value of spectral lag is considered to be positive in this case, as seen in fig.1c, left). So spectrum of the event evolves from hard to soft in that case. This type of spectral evolution is consistent with kinematic model (Ukwatta et al., 2012). Negative value indicates evolution of spectrum from soft to hard, and zero value means that there is no spectral evolution in the event. Negative and zero values are not consistent with kinematic model.

In 6 cases of 43 in the relation there is break (fig.1c, right). Two-logarithmic model (3.2-3.3) with exponential break is used for fitting the lag - energy relation of these events, where $E_{cut}$ is value of energy of the break. Parameter C determines sharpness of the break, and it's value is fixed to 100.

Spectral lag - energy relation for separate pulses of GRBs or for GRBs with simple structure of light curves (fig.1b, left) shows no break and no negative spectral lag (fig.1c, left). Break in the relation and negative lag may appear due to different spectral parameters of the overlapping pulses (fig.1c, right) and have no connection with physics of the GRB source.

For 9 well separated GRB pulses spectral lag index - duration relation was formed (fig.2). The relation is well fitted by power law with power index equal to $1.14\pm0.15$. There is one short GRB081226 in the sample and it does not violate the correlation. So short GRBs may follow the same law as the long ones and it may be the evidence of the same emission mechanism in short and long GRBs. This correlation was also found in paper (Hakkila and Preece, 2011). But in that paper spectral lag between light curves in BATSE energy channels was used as spectral evolution parameter instead of spectral lag index.

\section{Conclusions}
Spectral evolution of 28 bright GRBs registered by SPI and IBIS/ISGRI of INTEGRAL observatory was investigated.

It was found that for simple structure bursts or well separated pulses of multipulse bursts the spectral lag can be approximated by the relation of $ \tau \sim A\lg(E)$.

Parameter A (spectral lag index) is new alternative parameter characterizing spectral evolution of GRBs.

Spectral lag index correlates with pulse duration. The dependence of the spectral lag index on the duration of GRB pulses is presumably the same for long and for short GRBs.

Negative lag in simple structure bursts or in well separated pulses was not found.

\section{Acknowledgements}
The work was partially supported by RFBR grant 12-02-01336-a
\begin{figure}
\includegraphics[width=1.0\textwidth]{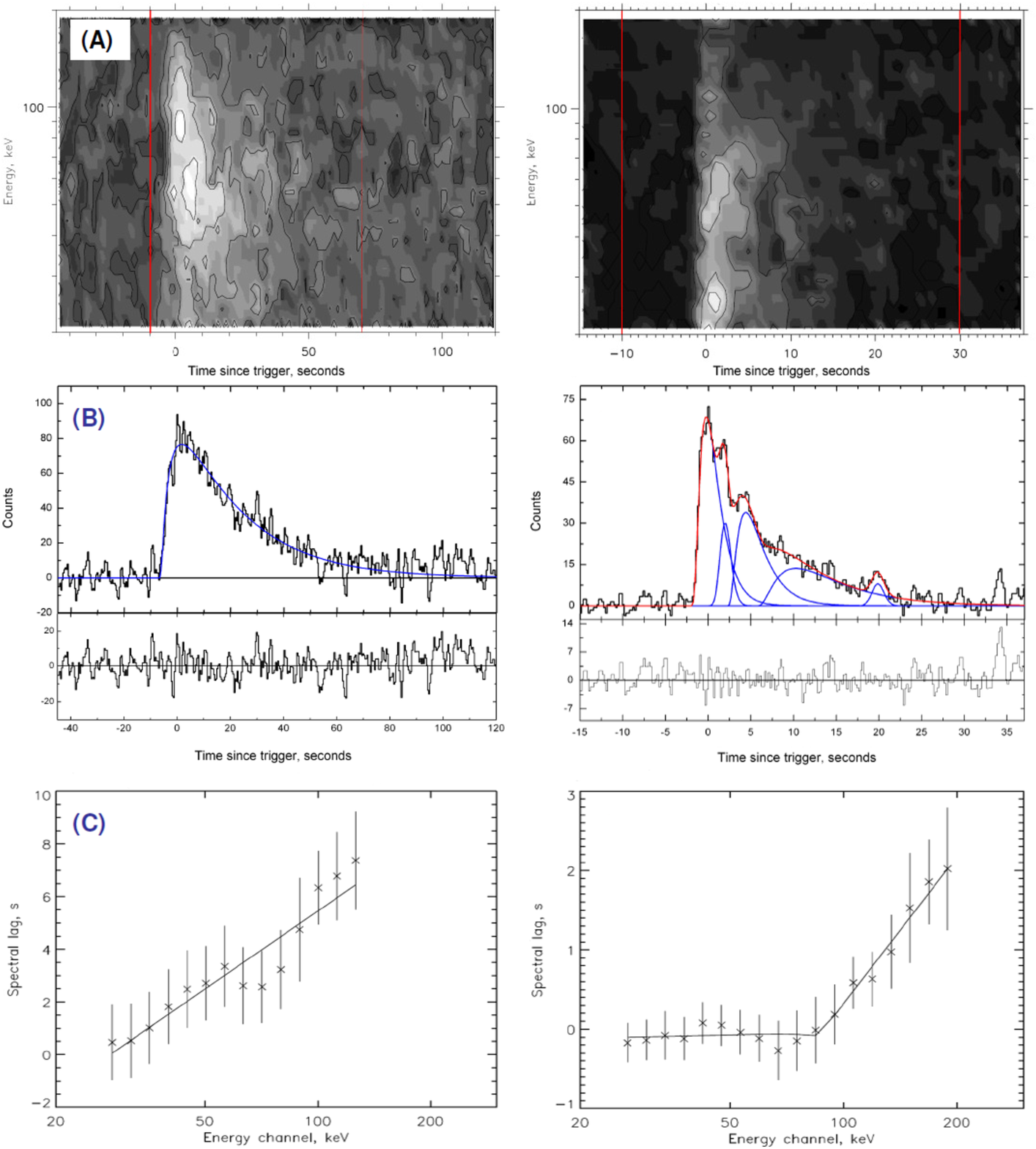}
\caption{GRB050504 (left) and GRB031203 (right). (A) - Energy - time diagram (IBIS/ISGRI data). Brightness of color in picture represents number of counts in energy-time bin (black color corresponds to minimal value). Red lines show time interval used in cross-correlation analysis.
(B) - At the top - mask weighted light curve in [20-200] keV energy band (IBIS/ISGRI data). Blue line represents 4-parametric fit of GRB pulse (Norris et al., 2005), red line represents total fit in case of GRB031203 (right) with multi-pulse structure. At the bottom - residuals of the fit.
(C) - Spectral lag - energy relation in [20-200] keV energy band (IBIS/ISGRI data). Errors are shown at 1$\sigma$ level. Black line represents fit of the curve with model (3.1) in case of GRB050504 (left) and model (3.2-3.3) for GRB031203 (right).}
\label{fig1}
\end{figure}

\begin{figure}
\includegraphics[width=1.0\textwidth]{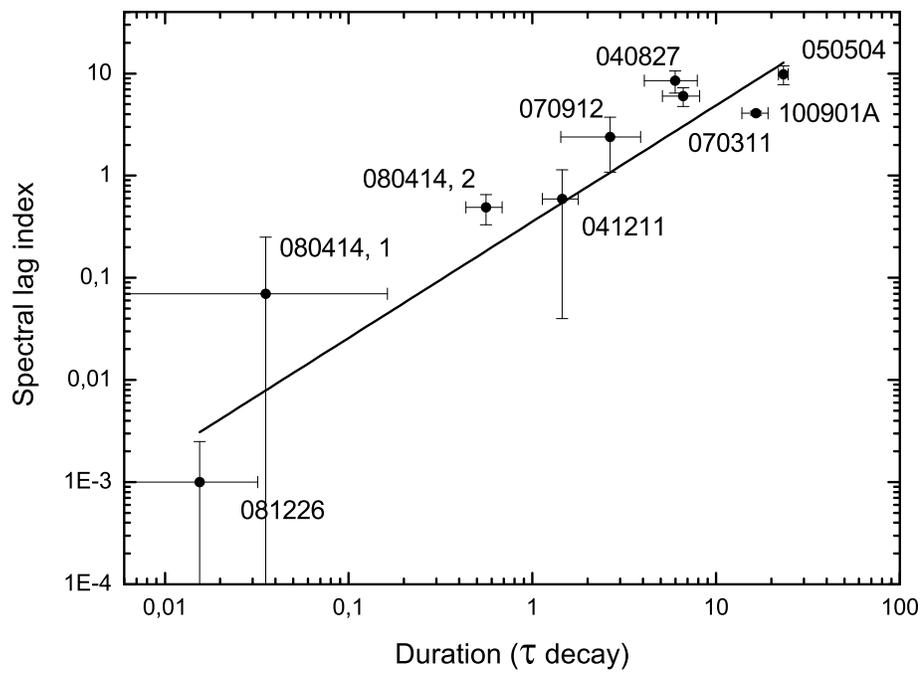}
\caption{The dependence of the spectral lag index on the duration of GRB pulses.}
\label{fig2}
\end{figure}

\end{document}